\documentclass[prb,pre,aps,twocolumn,amsmath,amssymb,floatfix,
superscriptaddress]{revtex4-1}
\usepackage[dvips]{graphics}
\usepackage{graphicx}
\usepackage{color}
\usepackage{blindtext}
\definecolor{dred}{rgb}{0.75,0,0}
\usepackage{graphicx}
\usepackage{dcolumn}
\usepackage{bm}
\usepackage{xcolor}
\usepackage{braket}
\usepackage{physics}
\usepackage{appendix}
\usepackage{hyperref}
\hypersetup{
 colorlinks=true,
citecolor=magenta,filecolor=blue,
 linkcolor=magenta,
 }

\usepackage{graphicx}
\usepackage{dcolumn}
\usepackage{bm}
\usepackage{xcolor}

\definecolor{codegreen}{rgb}{0,0.6,0}
\definecolor{codegray}{rgb}{0.5,0.5,0.5}
\definecolor{codepurple}{rgb}{0.58,0,0.82}
\definecolor{backcolour}{rgb}{0.95,0.95,0.92}

\begin{document}

\preprint{APS/123-QED}

\title{\textcolor{blue}{Tailoring flat bands and topological phases in a multi-strand Creutz network}} 

\author{Amrita Mukherjee}
\affiliation{Department of Physics, University of Kalyani, Kalyani,
West Bengal-741 235, India}
\email{amrita.physics@klyuniv.ac.in}

\author{Atanu Nandy}
\affiliation{Department of Physics, Acharya Prafulla Chandra College, New Barrackpore, Kolkata, West Bengal-700 131, India}
\email{atanunandy1989@gmail.com}

\author{Shreekantha Sil}
\affiliation{Department of Physics, Visva-Bharati, Santiniketan, 
West Bengal-731 235, India}
\email{shreekantha.sil@visva-bharati.ac.in}

\author{Arunava Chakrabarti}
\affiliation{Department of Physics, Presidency University, 86/1 College Street, Kolkata, West Bengal - 700 073, India}
\email{arunava.physics@presiuniv.ac.in}
\date{\today}

\begin{abstract}
We prove that, a suitable correlation between the system parameters can trigger topological phase transition and flat bands in a multi-strand  Creutz ladder network, when a staggered second neighbor interaction is included along the $x$-axis. An appropriate change of basis maps such a finite $N$-strand mesh into  $N$ or $N-1$ decoupled Su-Schrieffer-Heeger chains, depending on $N$ even or odd. A simple intuitive method, using a real space decimation scheme turns out to be very powerful in analytically extracting the flat bands, explaining their degeneracy or a lifting of the same. Our results are analytically exact, and may inspire experiments in photonics and ultracold atomic systems.

\end{abstract}

\maketitle

\section{Introduction}
\label{intro}
The occurrence of flat, non-dispersive energy bands in periodic crystalline lattice geometries~\cite{leykam1,flach,leykam2}, and a revelation of the marvels of topological phase transitions along with the concepts of symmetry protected edge states~\cite{asboth,su,heeger,zhao,dias1,dias2,dias3}, have been two important areas (among many) in condensed matter physics that have spurred immense research activity over a period of two decades now. 

The flatband (FB) networks, realized experimentally in the arena of photonics~\cite{seba1,rodrigo,seba2}, using ultrafast laser writing techniques, have unfolded a particularly interesting working platform where one encounters a class of perfectly localized compact modes, the so called `compact localized states' (CLS). The CLS's arise due to certain special geometries embedded in a perfectly periodic lattice, leading to a  destructive quantum interference between different connecting paths. The CLS's are the Hamiltonian eigenstates, and are marked by strictly vanishing amplitudes outside a spatial subcluster of the lattice points in a given geometry. In recent times, topological flat bands in decorated, Archimedian lattices in two dimensions have drawn considerable interest~\cite{biplab1,biplab2,crasto,yan}. The 
CLS's have also been identified as  prospective candidates for the storage and transfer of quantum information~\cite{rontgen}. This special geometry-induced localization may easily be differentiated from its counterpart, viz, the canonical case of a disorder-driven Anderson localization~\cite{anderson}.

The path breaking idea of a `topological' phase transition was put forward in the early 1970's~\cite{thouless1}. It was argued that a two dimensional system could exhibit a phase transition, engineered by the topological defects (vortices), that didn't break the Landau symmetry, as observed in the conventional phase transitions. Instead, it changed the {\it topology} of the system. The concept of the topological phases was further invoked to explain the remarkably precise quantization of the Hall conductivity in two dimensional electronic systems~\cite{thouless2}, and this made `topology' the arena of an intense research in condensed matter theory and experiments in the last couple of decades~\cite{dalibard}.

The two fields of interest, as described above, perfectly amalgamate in the analysis of electronic or photonic modes of excitations described on one, quasi-one or two dimensional decorated lattices. Experimental realization of flat bands with non-trivial topology in kagom\'{e} lattice structure~\cite{li} and, the Lieb architecture~\cite{xia,dias3,rodrigo} has been achieved in optical~\cite{taie}, and in ultracold atomic lattices~\cite{nathan,bloch}. These geometries have subsequently been analysed to unravel non-trivial localization properties~\cite{rudo1,rudo2}. 

The canonical example of a Su-Schrieffer-Heeger (SSH) chain~\cite{asboth,su}, that triggered the fundamental interest in the study of topological properties has also been an inspiration in the quest for a richer variety of topological properties in low dimensional lattice models with a nominal quasi-one dimensionality introduced in the systems. The simplest extensions have been the so called `coupled SSH chains'~\cite{padavic,sen,zhang}, and the preceding work on the chiral ladder models~\cite{hugel}. These quasi-one dimensional models generally consist of two strands, and in a way, simulate the  optical ladders loaded with cold atoms, systems that are now experimentally realizable in optical lattice-based experiments~\cite{aidelsburger}.
\begin{figure}[ht]
\centering
\includegraphics[clip,width=\columnwidth,angle=0]{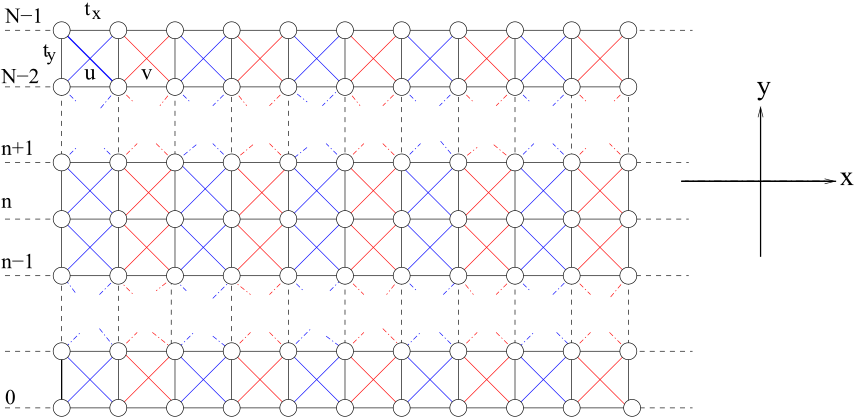}
\caption{(Color online) A quasi-one dimensional $N$-strand ladder network. The nearest neighbor hopping integrals along $x$ and $y$-directions are $t_x$ and $t_y$ respectively. The second neighbor hopping, along the diagonals, alternate between the values $u$ (blue) and $v$  (red), as shown.}
\label{ladder}
\end{figure}

The Creutz ladders~\cite{creutz} form a substantial part of the research focused on quasi-one dimensional ladder networks - chiral or non-chiral. Such topological ladders or their equivalents\cite{bermudez,cuadra,zurita} are realized experimentally using the laser assisted tunnelling,  and are now drawing serious attention from the experimentalists~\cite{aidelsburger,atala,junemann}. These systems have been really useful in unmasking the two dimensional topological phases and new symmetry classification schemes from their apparent one dimensional character~\cite{hugel}. Interestingly, the ladder networks belonging to the Creutz family have also been scrutinized in respect of the flat energy bands~\cite{orito}, their identity as a flat-band topological insulator~\cite{zurita}, or as a prototype system exhibiting the flat band many body localization~\cite{kuno}, to name a few (but definitely, not all) exciting recent works.

Motivated by the eruption of these developments, in the present communication we examine an hitherto unexamined (to the best of our knowledge) quasi-one dimensional network, namely, a Creutz strip network (CSN), in search of any flat bands and a topological phase transition. We choose the CSN to have an infinite extent along the $x$-direction, being restricted to a finite, but arbitrarily large number of strands $N$ in the $y$-direction. We introduce a staggered distribution of the second neighbor hopping along the $x$-axis, as shown in Fig. \ref{ladder}, making it belong to a class of an extended SSH model, where the staggering is introduced only in the `connectivity' among the distant neighbors.

We introduce a simple real space decimation scheme to discern the non-dispersive, flat energy bands occurring in such systems. The decimation scheme explains the degeneracy of the flat bands and also yields the condition needed to lift the degeneracy. Apart from exploring the flat bands, we explicitly work out the condition of observing the topological phase transition, based again on the prescribed real space decimation method. The topological invariant, viz, the Zak phase~\cite{zak} is calculated and the topologically protected edge states, that pay a tribute to the bulk-boundary correspondence~\cite{asboth}, are obtained and discussed.

In what follows we describe our findings. In section II we elaborate on how to extract the flat, non-dispersive bands in a CSN, as proposed here. The occurrence of degeneracy and its lifting is made clear by exploiting a real space decimation scheme that is simple and intuitively appealing. The cases of a two and a three strand CSN are explicitly discussed and the general $N$-strand case follows easily. Section III discusses the topological properties of the strip, namely, the topological invariant, the edge states and all that. In section IV we draw our conclusion.

\section{Discerning the Flat Bands}
Let's refer to Fig.~\ref{ladder} that represents an $N$-strand CSN, with an infinite extent along the $x$-axis.
The analysis begins with the tight binding Hamiltonian, 
\begin{equation}
    H = \epsilon \sum_j c_j^\dag ~ c_j + \sum_{j,k} [ t_{jk}~ c_j^\dag 
    c_k + h.c.]
    \label{ham}
\end{equation}
Here, $\epsilon$ is the `on-site' potential, taken a constant throughout. The hopping (overlap) integral $t_{jk}=t_x$ for hopping between the nearest neighbors along the $x$-axis. Along the $y$-axis, we assume hopping between the nearest neighboring strands only, and this hopping integral is designated by $t_y$. The staggered second neighbor interactions are along the diagonals, and alternate between the values $u$ (blue), and $v$ (red) along the $x$-axis. It should be noted that, such an $N$-strand mesh, in the presence of the second neighbor (diagonal) hopping integrals, makes the sites sitting on the edges, in the $y$-direction, to have a coordination number equal to four. The sites in the bulk of the system naturally, have eight nearest neighbors. Knowing the existence, if any, of the flat, non-dispersive bands, is thus a non-trivial task. Our scheme explains how the flat bands can occur, and for what choices of the parameters.
\begin{figure}[ht]
\centering
(a) \includegraphics[width=0.40\columnwidth]{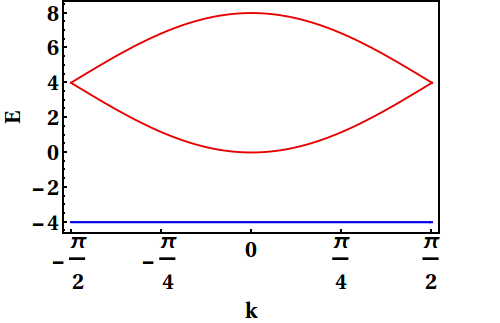}
(b) \includegraphics[width=0.40\columnwidth]{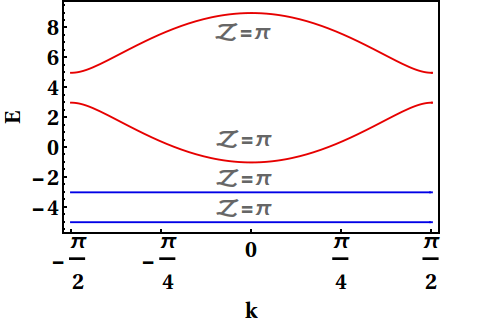}\\
(c) \includegraphics[width=0.36\columnwidth]{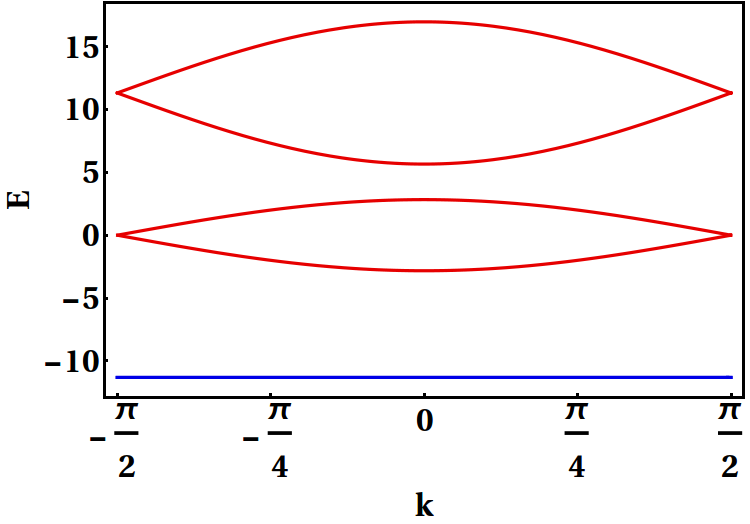}
(d) \includegraphics[width=0.41\columnwidth]{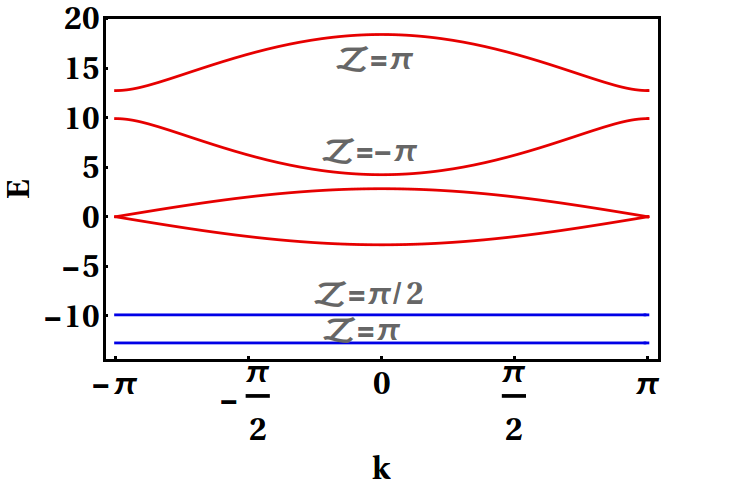}
\caption{(Color online) Energy-momentum dispersion relation for  \textit{two-strand} (a,b) and \textit{three-strand} (c,d) ladder networks. The values of the parameters are respectively (a) $t_x = 1$, $t_y = 4$, $u = 1 = v$ and (b) $t_x = 1$, $t_y = 4$, $u = 1$, $v = 2$, (c) $t_x = \sqrt{2}$, $t_y = 8$, $u= 1 = v$ and (d) $t_x = \sqrt{2}$, $t_y = 8$, $u = 1$, $ v = 2$.
For both the cases, when $u = v $ the flat band is doubly degenerate while the degeneracy is removed when $u \ne v$.}  
\label{ladderdisp}
\end{figure}

Let us now explain the basic working principle first in terms of a $2$-strand ladder with staggered second neighbor interactions, and then for a three-strand ladder.  In each case we restrict ourselves to only two kinds of second neighbor hoppings, though the scheme for a general $N$ with any number of staggered diagonal hopping integrals follow a similar analysis. Definitely these latter cases offer a richer in variety in terms of the flat bands (and topological properties as well) - their energies and degeneracies.

\subsection{Flat bands in a Creutz strip: degeneracy and it's lifting}
\subsubsection{ A two-strand ladder with staggered second neighbor hopping}
\begin{figure}[ht]
\centering
\includegraphics[width=\columnwidth]{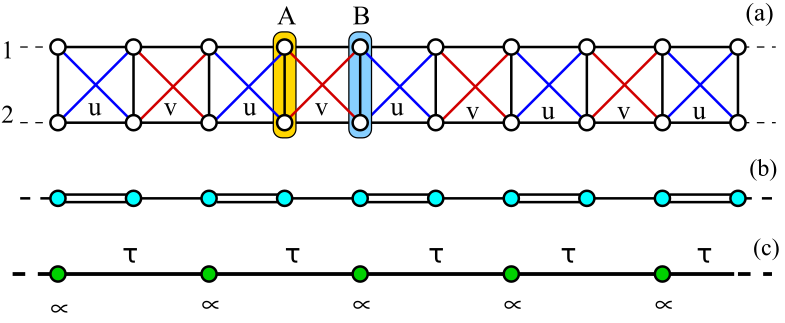}
\caption{(Color online) (a) A two-strand ladder network with staggered
second neighbor hoppings $u$ (blue) and $v$ (red), and the two sublattices $A$ and $B$. (b) The corresponding effective one dimensional chain (cyan colored sites) with the renormalized on-site potential $\epsilon-t_y$, and the hopping integrals $t_x-u$ (double bond) and $t_x-v$ (single bond). (c) The second decimation of the alternate sites in (b) maps it into a uniform periodic chain with an energy dependent on-site potential $\mu$ (the green colored sites) and a constant, energy dependent renormalized hopping integral $\tau$.} 
\label{ssh2strand}
\end{figure}

Fig.~\ref{ssh2strand}(a) shows a two-strand CSN with a periodic arrangement of second neighbor hopping (along the diagonals) integrals. We see that, there is a two-sublattice structure, $A$, and $B$ (shaded in golden and sky blue respectively), depending on the arrangement of the second neighbor hoppings on either side of $A$ and $B$. Sites on one sublattice, say $A$, have the diagonal connection (hopping) $u$ to their left, and $v$ to their right, and for the sites belonging to the $B$ sublattice, it is just the opposite. 

The time independent Schr\"{o}dinger equation can be cast in the equivalent `difference equation' forms, for any pair of sites $(j,1)$ and $(j,2)$ along the $j$-th rung of the ladder, as the following pairs of equations:
\begin{widetext}
\begin{eqnarray}
(E - \epsilon)\psi_{j,1} & = & t_x (\psi_{j+1,1} + \psi_{j-1,1} ) + t_y \psi_{j,2} + 
u \psi_{j-1,2} + v \psi_{j+1,2} \nonumber \\
(E - \epsilon)\psi_{j,2} & = & t_x (\psi_{j+1,2} + \psi_{j-1,2} ) + t_y \psi_{j,1} + 
u \psi_{j-1, 1} + v \psi_{j+1, 1}
\label{diff}
\end{eqnarray}
\end{widetext}
It is obvious that, if the site $j$ belongs to the 
$A (B)$-sublattice, then all its neighbors, nearest of the next nearest, belong to the sublattice $B (A)$.
Let us sequentially subtract and add the second equation in Eq.\eqref{diff} from (to) the first, and define $\phi_{j,1}=\psi_{j,1} - \psi_{j,2}$, and $\phi_{j,2}=\psi_{j,1} + \psi_{j,2}$  That is, we change the basis. Then from Eq.~\eqref{diff} we arrive at a simpler set of {\it decoupled} equations in this new basis, viz.  
\begin{eqnarray}
    [E-(\epsilon-t_y)]\phi_{j,1} & = & (t_x-u)\phi_{j-1,1} + (t_x-v) \phi_{j+1,1} \nonumber \\
    \left [ E - (\epsilon+t_y) \right ] \phi_{j,2} & = & (t_x+u) \phi_{j-1,2} + (t_x+v) \phi_{j+1,2} \nonumber \\
    \label{binary}
\end{eqnarray}
A look at the two equations in Eq.~\eqref{binary} immediately reveals that, the change of basis, from $\psi_j$ to $\phi_j$ casts the pair of original equations in Eq.~\eqref{diff} into a pair of equations, each representing a single, perfectly periodic chain with a staggered, alternating {\it effective} pair of hopping integrals $t_x - u$, and $t_x-v$ for the first, and $t_x + u$ and $t_x + v$ for the second. The effective on-site potentials read $\epsilon-t_y$, and $\epsilon + t_y$ respectively. Evidently, no physics is lost in this transformation. The original two-strand ladder is shown in Fig.~\ref{ssh2strand} (a), and the effective one dimensional chain, described in the new basis $\{\phi_{j,1} \equiv \psi_{j,1}-\psi_{j,2}\}, j \in [-\infty,\infty]$ is shown in Fig.~\eqref{ssh2strand} (b) respectively.

Let us, without loss of any generality, pick up the first of the two equations in Eq.~\eqref{binary}. This equation can further be renormalized, by eliminating the amplitudes $\phi_{j-1,1}$ and $\phi_{j+1,1}$ at alternate sites, so as as to have a difference equation representing a periodic lattice with a {\it single} on-site potential $\mu$, and a {\it uniform} hopping integral $\tau$, both being functions of the energy $E$. The {\it final} lattice is shown in Fig.~\ref{ssh2strand}(c). The difference equation describing this lattice reads, 

\begin{equation}
(E-\mu) \phi_{j,1} = \tau (\phi_{j-2,1} + \phi_{j+2,1})
\label{pure}
\end{equation}
with, 
\begin{eqnarray}
\mu & = & \epsilon - t_y + \frac{(t_x-u)^2 + (t_x-v)^2}{E-\epsilon+t_y} \nonumber \\
\tau & = & \frac{(t_x - u) (t_x - v)}{E-\epsilon+t_y}
\label{parameters}
\end{eqnarray}
The subscript $j$ in Eq.~\eqref{pure} runs over the the green colored sites in Fig.~\ref{ssh2strand}(c), that survive the decimation.

The set of Eqs.~\eqref{parameters} immediately reveals that, if we select either $t_x=u$, or $t_x=v$, the right hand site of Eq.~\eqref{pure} becomes zero and consequently, 
Eq.~\eqref{pure} admits a {\it localized} atomic-like state at an energy, which is a solution of the equation $E-\mu =0$, that is,  at $E=\epsilon - t_y \pm (v-u)$. This is a CLS, and gives a flat band. The energy eigenvalues can be easily verified from Fig.~\ref{ladderdisp}(a) and (b). The amplitudes of the wavefunction for such a FB can be worked out as well. One such configuration is $\psi_{j,1} = \pm \psi_{j,2} = 1$ (say) on sublattice $A$, and $\psi_{j,1} = \pm \psi_{j,2} =0$ on sublattice $B$, satisfying the Schr\"{o}dinger equation at the FB energy eigenvalues. Had we taken up the second equation of the set of Eq.~\eqref{binary}, the CLS would have appeared for $t_x = -u$ or $t_x = -v$, and at $E = \epsilon + t_y \pm (v+u)$.
The dispersion relations obtained from the pair of equations in Eq.~\eqref{binary} are easily obtained after scaling them once more, by decimating the alternate sites, as explained before. The result is, 
\begin{widetext}
\begin{equation}
    [E - (\epsilon \mp t_y)]^2 - [(t_x \mp u)^2 + (t_x \mp v)^2] - 2 (t_x \mp u) (t_x \mp v) \cos ka =0
    \label{disp}
\end{equation}
\end{widetext}
Here, $k$ is the wave vector, and $a$ is the lattice constant defined on the linear chain in Fig.~\ref{ssh2strand}(c). The sign `$-$' or `$+$' needs to be chosen depending on whether we are dealing with the first, or the second of the set of Eq.~\eqref{binary}.  

The non-dispersive bands can easily be traced back to the appropriate correlation between the hopping integrals.
It is easy to see that $t_x=u$, or $t_x=v$ (or, $t_x=-u$, or $t_x=-v$) makes the equation independent of the wave vector $k$ and the band becomes non-dispersive or flat. Also, the two-fold degeneracy for $t_x=u=v$ is obvious.
\begin{figure}[ht]
\centering
(a) \includegraphics[width=0.4\columnwidth]{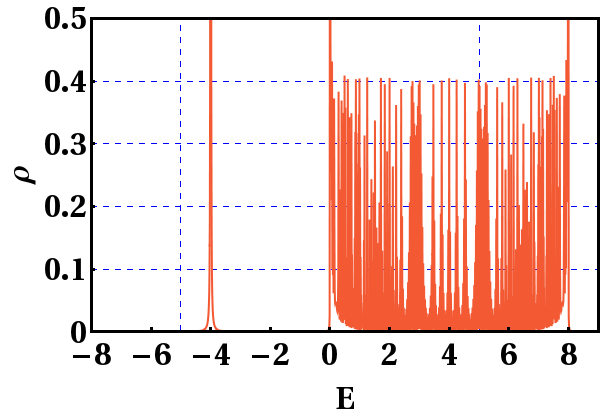}
(b) \includegraphics[width=0.4\columnwidth]{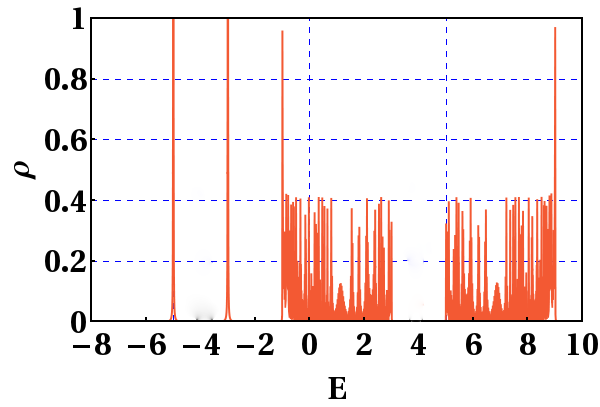}\\
(c) \includegraphics[width=0.40\columnwidth]{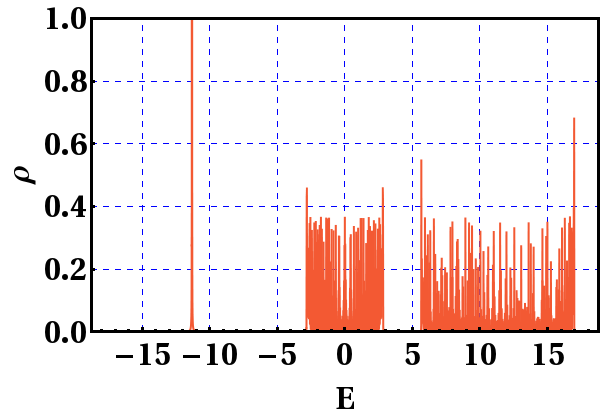}
(d) \includegraphics[width=0.40\columnwidth]{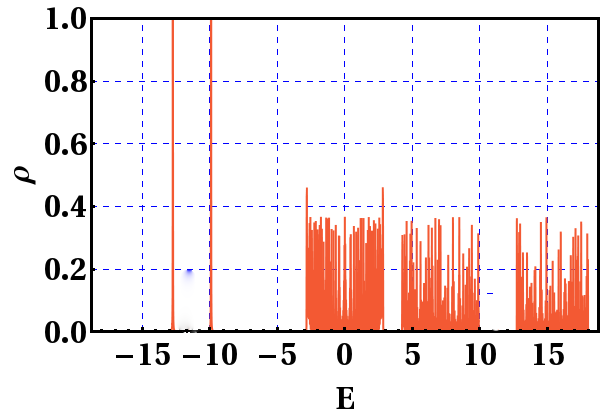}
\caption{(Color online) Variation of the density of states $(\rho)$ for a  \textit{two-strand} (a,b) and \textit{three-strand} (c,d)  CSN. The values of the parameters are respectively (a) $t_x = 1$, $t_y = 4$, $u = 1 = v$ and (b) $t_x = 1$, $t_y = 4$, $u = 1$, $v= 2$.
 The degenerate CLS at $E=-4$ for $u = v$ is
 seen to be present. The degeneracy is lifted for $u \ne v$, and the FB states appear at $E=-3$ ,$-5$. The topologically protected edge state lying in between, is discussed in text. In (c) we show the overall DOS and the two-fold degenerate FB for $t_x=\sqrt{2}$, $t_y=8$, $u=v=1$, and in (d) we select $t_x=\sqrt{2}$, $t_y=8$, $u=1$, and $v=2$, to show the removal of the degeneracy of the FB's.} 
\label{dos}
\end{figure}

Let us specifically talk in respect of the basis $\{ \phi_{j,1} \equiv \psi_{j,1} - \psi_{j,2} \}, j \in [-\infty,\infty]$.
 For $v \ne u$, the degeneracy is lifted, but as we choose $t_x=u$ or $v$, the non-dispersive bands still show up at $E=-t_y \pm (v-u)$. The results tally exactly with Fig.~\ref{ladderdisp}, where, with $t_y=4$, $u=2$, and $v=1$, the flat bands appear at $E=-5$, and at $E=-3$ (Fig.~\ref{ladderdisp}(b)), while a doubly degenerate flat band appears for $u=v=1$ at $E=-t_y=-4$ (Fig.~\ref{ladderdisp}(a)). Energy $E$ is always measured in unit of $t_x$. We have chosen $\epsilon=0$. The two-fold degenerate CLS shows up at $E=-4$ in the display of the density of states (DOS) of a Creutz strip, and at $E=-3$ and at $E=-5$ when the degeneracy is lifted. Fig.~\ref{dos}(a) and (b) depict these cases respectively. 
 
The atomic-like CLS, that represents the FB is clearly visible in the density of states (DOS) spectrum shown in Fig.~\ref{dos}(a). The DOS is calculated using a Green's function formalism for a finite sized system. With the parameters chosen for the discussion so far, the DOS in Fig.~\ref{dos}(a) exhibits a sharp delta-like spike at $E=-4$, which is exactly the FB energy depicted in Fig.~\ref{ladderdisp}(a). The localized FB state appears outside the main band of extended states ranging between $E = [0,8]$ for the selection of the parameters elaborated in the text. With $u \ne v$ the degenerate FB in (a) splits into two, at $E=-3$ and $E=-5$, as we have already proved. There is a smaller localized peak in between these two FB states that corresponds to a topological edge state, as we will see later.

Needless to say, similar arguments can be woven if we use the basis 
$\{ \phi_{j,2} \equiv \psi_{j,1} + \psi_{j,2} \}, j \in [-\infty,\infty]$. In this case we would need to set $t_x=-u$ or $t_x=-v$. The negative sign in front of the hopping integrals do not affect the energy spectrum. The flat band now appears at $E=t_y$. The degeneracy and its removal follow the same line of arguments. However, we have not shown this set of images here, to save space.

Before we end this subsection, an obvious and a very pertinent discussion regarding Eq.~\eqref{pure} is in order. A situation with $\tau=0$ on the the right hand side of Eq.~\eqref{pure} is also satisfied by setting $E \ne \mu$, but with $\phi_{j,1} \equiv  \psi_{j,1} - \psi_{j,2} = 0$. This implies $\psi_{j,1} = \psi_{j,2}$ for all $j$. We can easily exploit this information to work out the band of the allowed energy eigenvalues given by the second of the Eq.~\eqref{binary}. A simple algebra shows that the band edges obtained from the second equation in Eq.~\eqref{binary} are given by, 
\begin{equation} 
E = t_y \pm \sqrt{9 u^2 + v^2 + 6 u v}
\label{edges}
\end{equation}
when we set $t_x=u$, for example.
With $t_y = 4$, and $t_x=u=v=1$ the edges of the energy band turn out to be at $E=0$ and at $E=8$. These edges, as we mentioned just now, are obtained from the second of Eq.~\eqref{binary}. This is important, as we need to consider both the equations in Eq.~\eqref{binary} to get the full spectrum. The edge energy values exactly match with the DOS displayed in Fig.~\ref{dos}(a). There is a continuous distribution of the energy eigenvalues, all corresponding to Bloch-like extended states within this band.
\begin{figure}[ht]
\centering
\includegraphics[width=\columnwidth]{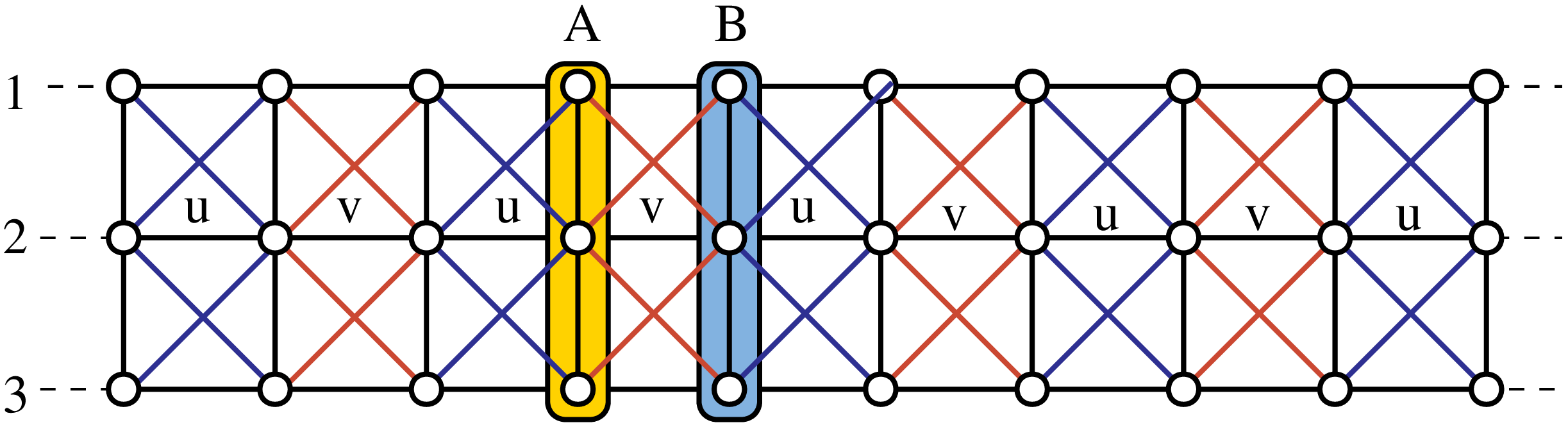}
\caption{(Color online) The three-strand Creutz strip with infinite 
extent along $x$-direction. The $A$ and $B$ sublattices are highlighted.} 
\label{ssh3strand}
\end{figure}
\subsubsection{The three-strand ladder}
The power of the proposed change of basis, resulting in a decoupling of the 
coupled difference equations for a multi-strand ladder, to discern the flat bands,  is appreciated further if we consider a three-strand ladder, as shown in Fig.~\ref{ssh3strand}. The strands are marked as `$1$', `$2$' and `$3$', and the sublattices $A$ and $B$ are painted in golden and blue, just like before. Since we have three strands now, its advantageous to cast the difference equation in a matrix form, connecting the amplitudes of the wavefunction at the three sites along any vertical $j$-th strand on an $A$ (or, a $B$) sublattices. The matrix formalism was used earlier to study a quasiperiodic ladder network~\cite{shreekantha}, and a two dimensional system with correlated disorder~\cite{alberto}, and is used in the same spirit here.

Let's define, for any sublattice $A$ or $B$, 
\begin{equation}
\bf \Psi_{j} \equiv  
\left[ \begin{array}{c}
\psi_{j,1} \\
\psi_{j,2} \\
\psi_{j,3}  
\end{array}
\right ] 
\end{equation}
where $j$ denotes the site index, and `1', `2' and `3' give the `strand index'.
The difference equation, in matrix form, now appears to be,   
\begin{eqnarray}
(E\bf I - \tilde {\bf \epsilon}) \bf \Psi_{j,A (B)} = 
\tilde{\bf t}_{L} \bf \Psi_{j-1, B (A)} + 
\tilde{\bf t}_{R} \bf \Psi_{j+1, B (A)}
\label{diff3strand}
\end{eqnarray}
where, {\bf I} is the $3 \times 3$ unit matrix, $\tilde {\bf {\epsilon}}$ is the `potential matrix' given by, 
\begin{equation}
\mathcal{\tilde \epsilon} = \left[ \begin{array}{cccccccccccccccc}
\epsilon & t_y & 0  \\
t_y & \epsilon & t_y \\
0 & t_y & \epsilon 
\end{array}
\right ] 
\label{pot}
\end{equation}
The hopping matrices $\tilde{\bf t}_{L (R)}$ connecting the $j$-th rung on an $A$ sublattice to the $j-1$-th and the $j+1$-th ones on the $B$ sublattice on its left and right sides,
\begin{equation}
\tilde{\bf t}_L = \left[ \begin{array}{cccccccccccccccc}
t_x & u & 0  \\
u & t_x & u \\
0 & u & t_x 
\end{array}
\right ] 
~; 
\tilde{\bf t}_R = \left[ \begin{array}{cccccccccccccccc}
t_x & v & 0  \\
v & t_x & v \\
0 & v & t_x 
\end{array}
\right ] 
\label{hop}
\end{equation}
Obviously, while writing the hopping matrices for the $B$ rung one needs to replace $u (v)$ by $v (u) $ in $\tilde{\bf t}_{L (R)}$.

It is immediately verified that, the commutators $[\mathcal{\tilde \epsilon}, \tilde{\bf t}_L] = [\mathcal{\tilde \epsilon}, \tilde{\bf t}_R] = 0$ irrespective of the energy $E$. This means that, all the three matrices are simultaneously diagonalizable using the same matrix $\mathcal M$ (say). Taking advantage of this, we make a change of basis, going from the $\{\bf \Psi_j \}$ basis, to a new basis $\{ \bf \Phi_j \} = \mathcal M^{-1} \{ \bf \Psi_j \}$, which, when written explicitly reads, 
\begin{equation}
\left [ \begin{array}{c}
\phi_{j,1} \\ \phi_{j,2} \\ \phi_{j,3} \end{array} \right ] = 
\mathcal M^{-1} \left[ \begin{array}{c}
\psi_{j,1} \\
\psi_{j,2} \\
\psi_{j,3}  
\end{array}
\right ] 
\end{equation}
In the new basis, we now have three decoupled equations corresponding to the strands 1, 2 and 3, viz, 
\begin{widetext}
\begin{eqnarray} 
(E-\epsilon)~ \phi_{j,1} & = & t_x~(\phi_{j-1,1} + \phi_{j+1,1})  \nonumber \\
\left [E-(\epsilon-\sqrt{2} t_y) \right ] \phi_{j,2} & = & (t_x-\sqrt{2} u) ~\phi_{j-1,2} + (t_x-\sqrt{2} v)~ \phi_{j+1,2} \nonumber \\
\left [E-(\epsilon+\sqrt{2} t_y) \right ] \phi_{j,3} & = & (t_x+\sqrt{2} u) ~\phi_{j-1,3} + (t_x+\sqrt{2} v) ~\phi_{j+1,3} 
\label{diff3}
\end{eqnarray}
\end{widetext}
It's now easy, in the spirit of the discussion, to analyze the existence of the FB's and check their dispersionless character using the decimation argument used for the two-strand network case. For example, in the first of the set of Eqs.~\eqref{diff3}, setting $t_x=0$ created a FB at $E=\epsilon$, irrespective of the choice of the second neighbor hoppings $u$ and $v$.  This FB at $E=\epsilon$ will be present in all Creutz strips with an odd number of strands. The three-strand CSN in this case becomes an extended version of the simple diamond network used recently, to study the influence of a magnetic field on the FB's and the topological phases~\cite{amrita}.

Similarly, from the second and the third of Eqs.~\eqref{diff3}, setting $t_x=\pm \sqrt{2}u=\pm \sqrt{2}v$ yield doubly degenerate FB's at $E=\epsilon-\sqrt{2}t_y$ and $E=\epsilon+\sqrt{2}t_y$ respectively. Making $u \ne v$ lifts the degeneracy. 
The threads of the arguments follow exactly the same line as discussed before, and need not be repeated here. The discussion above is corroborated by the DOS of a three strand Creutz ladder, as shown in Fig.~\ref{dos}(c) and (d).
\begin{figure}[ht]
\centering
(a) \includegraphics[width=\columnwidth]{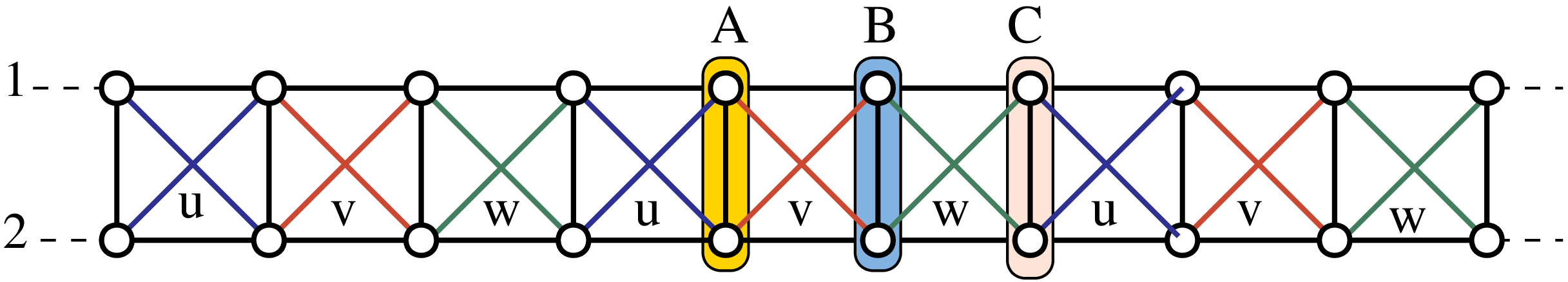}\\
(b) \includegraphics[width=0.50\columnwidth]{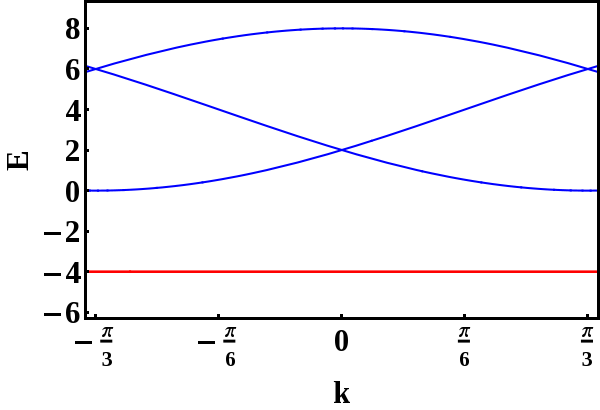}\\
(c) \includegraphics[width=0.50\columnwidth]{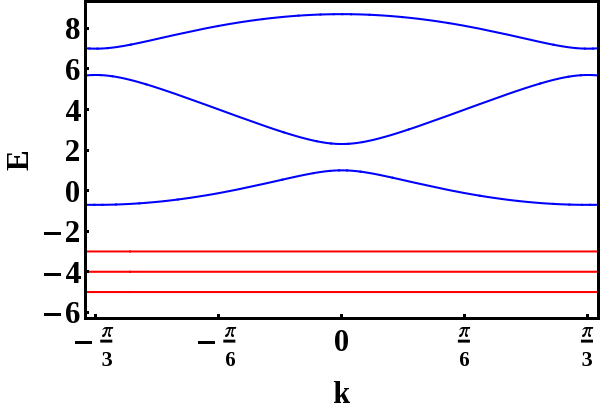}
\caption{(Color online) (a) A two-strand Creutz strip with a periodic repetition of three kinds of second neighbor (diagonal) hoppings $u$ (blue), $v$ (red) and $w$ (green). (b) The dispersion relation when $u=v \ne w$, revealing a two-fold degenerate flat band. (c) The removal of the degeneracy on making $u = w \ne v$
The values of the parameters are (b) $t_x = 1$, $t_y = 4$, $u = v = 1 = w$ and (c)
$t_x = 1$, $t_y = 4$, $u = 1 = w$, $v = 2$ respectively.}  
\label{tertiary}
\end{figure}

Before we leave this section, its relevant to draw the attention of the reader to the fact that, the scheme laid out above works perfectly well to unravel the FB's and their degeneracies or the lifting of it, in an $N$-strand ladder network with {\it any number} of second neighbor interactions, arranged periodically. The same decoupling scheme, initiated by an appropriate change of basis will conveniently extract all the different combinations of the intra and inter-strand hopping integrals, need to enforce FB's in the energy spectra. The topological issues, that are discussed later, can also be understood in the same way as we do for the two, and the three strand ladder here. Though we decide not to go into such details, for the sake of completeness, we present an energy-momentum dispersion relation of a two-strand Creutz ladder with three kinds of second neighbor (diagonal hoppings) interactions $u$, $v$ and $w$ arranged periodically along the $x$ direction. The lattice and the dispersion relations are presented in Fig.~\ref{tertiary}. The degenerate FB appears in (a), while in (b) we see that the degeneracy is lifted just by making $u \ne v$.  

\section{Topological properties}
Let us refer to the two-strand ladder network. The change of basis decouples the two strands, and each member of the set of two equations in Eq.~\eqref{binary} unfolds two different SSH chains with the staggered pairs of overlap integrals $(t_x-u, t_x-v)$ and $(t_x+u, t_x+v)$ respectively. Choosing any one of the equations, and setting say, $t_x - u > t_x-v$ or $t_x - u < t_x-v$ takes the system from one topological phase to another.

\subsection{The Topological invariant}
Let us get back to Fig.~\ref{ladderdisp} (b) and (d). To inspect the topological invariant, its quantization and the consequential topological phase transition, we 
write the Hamiltonian for a two-strand Creutz ladder in reciprocal space (k-space), viz, 
\begin{widetext}
\begin{equation}
\hat{\mathcal{H}}(k) = \left[ \begin{array}{cccccccccccccccc}
0 & t_y & t_{x} (1+e^{-2ika}) & u + v e^{-2ika} \\
t_y & 0 & u + v e^{-2ika} & t_{x} (1+e^{-2ika}) \\
t_{x} (1+e^{2ika}) & u +v e^{2ika} & 0 & t_y \\
u +v e^{2ika} & t_{x} (1+e^{2ika}) & t_y & 0
\end{array}
\right ] 
\label{hamilton}
\end{equation}
\end{widetext}
A similar Hamiltonian, but now of dimension $6 \times 6$ can easily be written down for a three-strand ladder. However, with increasing number of strands the matrix becomes more and more intricate, and a clear understanding of the dispersion relations and the topological aspects are conveniently obtained through the use of the decoupling scheme and the decimation technique. Here we discuss in terms of the two-strand case only, to save space.

 We set the lattice constant $a=1$, and $\epsilon=0$. It is seen here that, the matrix $\hat{\mathcal{H}}(k)$, is, by construction, time-reversal symmetric obeying $\hat{\mathcal{H}}(-k)^\ast = \hat{\mathcal{H}}(k)$, and also exhibits chiral symmetry. The chiral symmetry operator in this case is, 
\begin{equation}
\hat{\Gamma} = \left[ 
\begin{array}{c|c} 
  {\bf {\sigma_z}} & \mathcal{O} \\ 
  \hline 
  \mathcal{O} & {\bf {\sigma_z}}
\end{array} 
\right] 
\label{chiral}
\end{equation}
where, $\bf{\sigma_z}$ is the Pauli matrix, and $\mathcal{O}$ represents a $2 \times 2$ null matrix. It is easily verified that, $\hat{\Gamma}~\hat{\mathcal{H}}(k)~\hat{\Gamma}^{\dag} = - \hat{\mathcal{H}}(k)$, the basic requirement for the system to exhibit chiral symmetry.

This form of the Hamiltonian is also used to cross-check the dispersion relation and the FB's, that were initially obtained through decimation technique, and a decoupling of the coupled difference equations.

Making $u \ne v$ makes the gap open up both for the two, and the three strand ladders (and subsequently, for an $N$-strand ladder). 
The opening of an energy gap that was closed for $u=v$ at the Brillouin zone boundary is indicative of a topological phase transition.  We expect a topological invariant
to be associated with this phenomenon, a quantity such as the Zak phase~\cite{zak} that flips its quantized value from {\it unity} (in unit of $\pi$) to {\it zero} corresponding to the non-trivial and the trivial insulating phases respectively.  Recent experiments have suggested mechanisms for a possible measurement of this topological invariant~\cite{atala}.

The Zak phase for the $n$-th bulk bands is defined as,
\begin{equation}
    Z = \oint_{BZ}  \mathcal{A}_{nk}(k) dk
    \label{zak} 
\end{equation}
where $\mathcal{A}_{nk}$ is called the Berry  curvature of the $n$-th Bloch eigenstate, which is again defined as,
\begin{equation}
    \mathcal{A}_{n{k}}(k)= \bra{\psi_{n{k}}}\ket{\frac{d\psi_{nk}}{dk}}
\end{equation}

The integral is performed around a closed loop in the Brillouin zone, and $\ket{\psi_{nk}}$ is the $n$-th Bloch state. We use the Wilson loop approach~\cite{fukui}, a gauge invariant formalism. It protects the numerical value of the Zak phase against any arbitrary phase change of Bloch wavefunction. In Fig.~\ref{ladderdisp}(b) the bands exhibit an opening of the energy gap at the Brillouin zone boundary, and for $u<v$ the Zak phase assumes a quantized value of $\pi$, including the two non-degenerate FB's. With $u>v$ the Zak phase is identically zero for every band, confirming a topological phase transition in the Creutz ladders. Incidentally, it is observed that, experimental realization of topological FB's in a frustrated kagom\'{e} metal has recently been reported~\cite{kang}. A theoretical analysis of {\it nearly} flat bands with non-trivial topology has previously been pointed out in the literature~\cite{sankar}. The present geometry, simpler in nature may be inspiring to experimentalists, to test the existence of topological flat bands in a rectangular mesh.

\subsection{The edge states}

\begin{figure}[ht]
(a)\includegraphics[width=0.65\columnwidth]{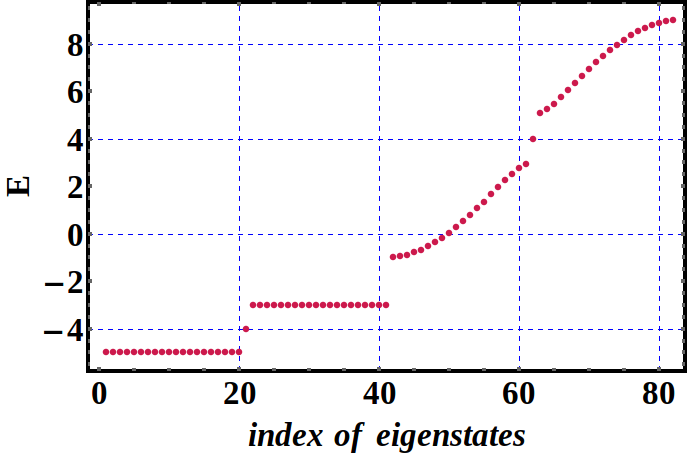}\\
(b)\includegraphics[width=0.65\columnwidth]{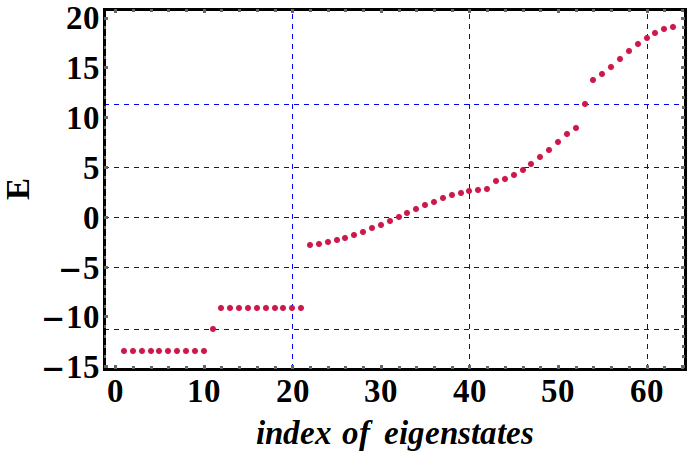}
\caption{(Color online) 
Distribution of energy spectra for a two arm and three arm Creutz strip network with open boundary condition for $N=20$ unit cells. The values of the parameters are chosen as (a) $t_x = 1$, $t_y = 4$, $u = 1$, $v = 2$. (b) $t_x = \sqrt{2}$, $t_y = 8$, $u = 1$, $v = 2.5$. The pairs of edge states are at (a) $\pm 4$, and (b) $\pm \sqrt{2} t_y$ and placed on blue dashed lines.}
\label{edge-twoarm}
\end{figure}

From the pair of Eqs.~\eqref{binary}, each one of which represents an SSH chain with $u \ne v$, it is easy to work out the chiral symmetry protected edge states when one has a finite array of the unit cells. The analyses go exactly parallel to what we already do for a normal SSH model, and need not be repeated here. However, the results, as depicted in Fig.~\ref{edge-twoarm} for a two-strand and a three-strand CSN may be elaborated a bit. For example, in case of a two-strand CSN, from the first of the Eqs.~\eqref{binary}, representing an effective SSH chain in the basis $\phi_{j,1} = \psi_{j,1} - \psi_{j,2}$, an edge state exists at $E= \epsilon - t_y$ if one deals with a finite segment of such an {\it effective} SSH chain with an integer number of unit cells. With $t_x=u \ne v$, this state is topologically non-trivial, as the corresponding Zak phase is quantized in units of $\pi$ (the bulk-boundary correspondence). This is exactly what we see when we deal with the full two-strand Creutz network, represented by the Hamiltonian in Eq.~\eqref{hamilton}. With $\epsilon=0$, and $t_y=4$, the topologically protected edge state appears at $E=-4$. Similarly, its other variant, obtained from the second of the Eqs.~\eqref{binary}, is at $E=4$.

\begin{figure}[ht]
(a)\includegraphics[width=0.40\columnwidth]{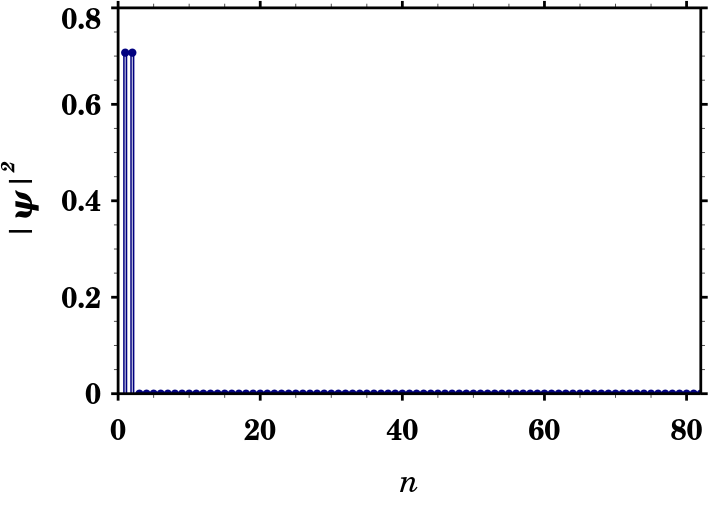}
(b)\includegraphics[width=0.40\columnwidth]{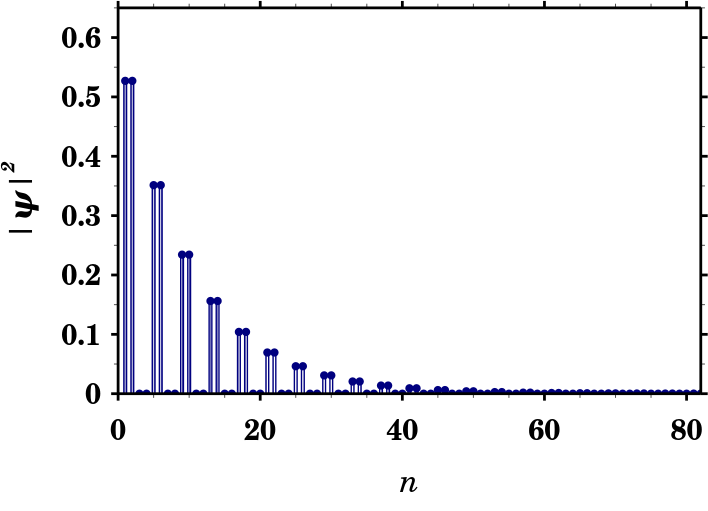}\\
(c)\includegraphics[width=0.40\columnwidth]{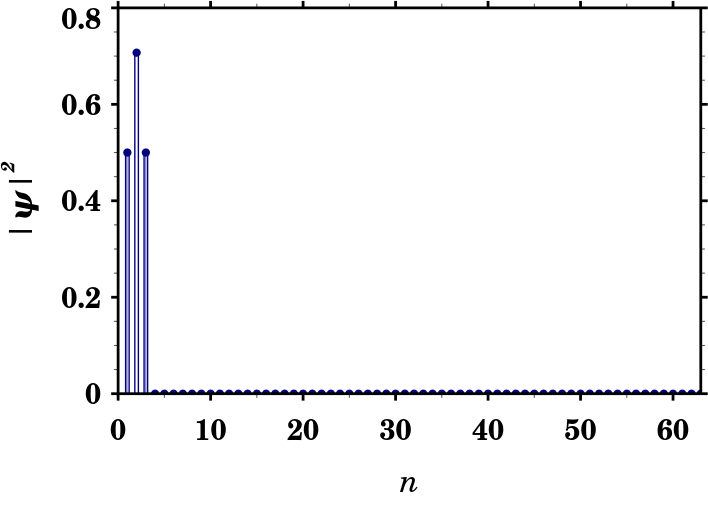}
(d)\includegraphics[width=0.40\columnwidth]{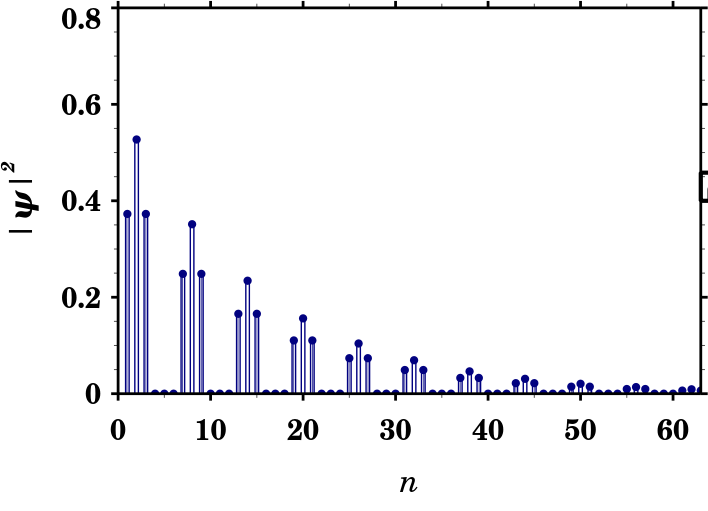}\
\caption{(Color online) 
Amplitude distributions of two pairs of in-gap edge states with energies $E = \pm 4$ and $E = \pm \sqrt{2} t_y$ for a \textit{two-strand} and a \textit{three-strand} ladder network respectively with open boundary condition for $N=20$ unit cells. The values of the parameters are (a,b) $t_x = 1$, $t_y = 4$, $u = 1$, $v = 2$ and (c,d) $t_x = \sqrt{2}$, $t_y = 8$, $u = 1$, $v = 2$.  Here the wavefunctions for energies (a) $E = -t_y$ and (c) $E = -\sqrt{2} t_y$ are sharply confined only at the left rung of both ladder networks. (b,d) portray a decay of wave amplitudes from the left edge of both ladder networks respectively for the energy (b) $E= t_y$ (two-strand case) and (d) $E= \sqrt{2} t_y$ (three-strand case).}
\label{edgeampli-twoarm}
\end{figure}

The edge states for a 3-strand ladder appear when we set $t_x = \sqrt{2} u$. With $u=1$, the state appears at $E=\sqrt{2} t_y$, as is seen in Fig.~\ref{edge-twoarm}(b), where we have set $t_y=8$. The distribution of the amplitudes in both the two-strand and the three-strand cases are shown in Fig.~\ref{edgeampli-twoarm}. The edge states at $E=-t_y$ (for the two-strand CSN, in Fig.~\ref{edgeampli-twoarm}(a)) and at $E=- \sqrt{2} t_y$ (for the three-strand CSN, in Fig.~\ref{edgeampli-twoarm}(c)) are sharply localized around the left edges of the sample, while the states with $E=t_y$ (Fig.~\ref{edgeampli-twoarm}(b)) and $E=\sqrt{2}t_y$ (Fig.~\ref{edgeampli-twoarm}(d)) corresponding to the two- and the three-strands CSN's exhibit an exponentially decaying penetration into the bulk of the system.

\section{Conclusion}
We have shown how a suitable change of basis can turn out to be instrumental in bringing out the flat band states in an $N$-strand Creutz ladder. A simple real space decimation scheme enables us to determine the criteria for degeneracy, or it's removal. The second neighbor staggered hopping, be it a binary or a higher order staggering, unravels the topological features of the system. This latter observation is facilitated again by the change of basis technique, that unveils the hidden SSH character in the Creutz strip. The topological invariant and the edge states have been explicitly obtained. The present day advancement in designing tailor-made optical and cold atom systems is inspiring and the present analysis may be put to test and be used to suggest possible applications.


\section{Acknowledgments}
A. M. acknowledges DST for providing her INSPIRE Fellowship $[IF160437]$. Both A. M. and A. N. thank Presidency University for providing the computational facility.


\end{document}